\documentclass[a4paper,twocolumn,11pt,accepted=2023-09-07]{quantumarticle}
\pdfoutput=1
\usepackage{amsmath,verbatim,latexsym,amssymb,indentfirst,mathrsfs,mathtools,amsthm,bbm,bm,hyperref,url,cancel,subcaption,enumitem}
\usepackage[font=small,labelfont=bf,format=plain,justification=raggedright,singlelinecheck=false]{caption}
\usepackage{verbatim,indentfirst}
\usepackage[title,titletoc]{appendix}
\usepackage{color}
\usepackage[dvipsnames]{xcolor}
\usepackage[normalem]{ulem}
\usepackage{physics}

\usepackage{graphicx}
\usepackage{placeins}
\usepackage{dcolumn}
\usepackage{bm}
\usepackage{multirow}

\begin{document}

\title{Experimental Communication Through Superposition of Quantum Channels}

\author{Arthur O. T. Pang}
\email{arthur.pang@mail.utoronto.ca}
\author{Noah Lupu-Gladstein}
\email{nlupugla@physics.utoronto.ca}
\author{Hugo Ferretti}
\email{hferrett@physics.utoronto.ca}
\author{Y. Batuhan Yilmaz}
\email{ybylmaz@physics.utoronto.ca.}

\affiliation{Department of Physics  and  Centre for Quantum Information   Quantum Control University of Toronto$,$ 60 St George St$,$ Toronto$,$ Ontario$,$ M5S 1A7$,$ Canada}

\author{Aharon Brodutch}
\email{brodutch@physics.utoronto.ca}
\affiliation{Department of Physics  and  Centre for Quantum Information   Quantum Control University of Toronto$,$ 60 St George St$,$ Toronto$,$ Ontario$,$ M5S 1A7$,$ Canada}
\affiliation{IonQ Canada Inc. 2300 Yonge St, Toronto ON, M4P 1E4}

\author{Aephraim M. Steinberg}
\email{steinberg@physics.utoronto.ca}

\affiliation{Department of Physics  and  Centre for Quantum Information   Quantum Control University of Toronto$,$ 60 St George St$,$ Toronto$,$ Ontario$,$ M5S 1A7$,$ Canada}
\affiliation{Canadian Institute for Advanced Research$,$ Toronto$,$ Ontario$,$ M5G 1M1$,$ Canada}
\date{2023-09-18}

\begin{abstract}
Information capacity enhancement through the coherent control of channels has attracted much attention of late, with work exploring the effect of coherent control of channel causal orders, channel superpositions, and information encoding. Coherently controlling channels necessitates a non-trivial expansion of the channel description, which for superposing qubit channels, is equivalent to expanding the channel to act on qutrits. Here we explore the nature of this capacity enhancement for the superposition of channels by comparing the maximum coherent information through depolarizing qubit channels and relevant superposed and qutrit channels. We show that the expanded qutrit channel description in itself is sufficient to explain the capacity enhancement without any use of superposition.
\end{abstract}

\maketitle


\section{\label{sec:intro}Introduction}
In a surprising paper, Ebler \textit{et al.} \cite{Ebler2018} showed that when two fully depolarizing channels are put in a superposition of causal orders \cite{Chiribella2013}, information can be transmitted. 
In their scheme, the order of two depolarizing channels which act sequentially on a system state is conditioned on a control qubit's state in the computational basis (states $|\textbf{a}\rangle$ and $|\textbf{b}\rangle$ in this paper), and by preparing and post-selecting the control qubit in $\left|+\right>\equiv\left(\left|\textbf{a}\right>+\left|\textbf{b}\right>\right)/\sqrt{2}$, information encoded on the system state can be transmitted.
This has been experimentally demonstrated in \cite{Taddei2021,Goswami2020,Rubino2017,Guo2020,Procopio2015,Rubino2021} and further theoretically explored in \cite{Procopio2019,Procopio2020,Procopio2020,Chiribella2019,Chiribella2021,Chiribella2021cyc,Sazim2021}.

This capacity enhancement phenomenon is, however, not unique to channels being placed in an indefinite causal order, and a similar capacity enhancement phenomenon has also been demonstrated by \cite{Gisin2005,Oi2003,Abbott2020,Guerin2019,Chiribella2019,Massa2019,DelSanto2018}. Notably, all of these schemes involve coherent control of the noise channel through the use of an ancillary qubit, where depending on the control state, the system state experiences indefinite ordering of noise channel, superposition of information encoding, or superposition of channels \cite{Rubino2021}. 

Here, we experimentally explore the nature of the capacity enhancement achieved by superposing two independent noisy qubit channels. As in the cases of the previously demonstrated schemes involving coherent control of noise channels, the superposition of two channels necessitates an expanded description of those channels to account for this extra control state. For the case of superposing qubit channels, this expanded description takes the form of a qutrit channel, as has been discussed briefly in the past \cite{Rubino2021,Abbott2020,Guerin2019,Chiribella2019}.

Expanding the channel description in this fashion is generally non-trivial, and the expansion is also not unique without additional information\cite{Araujo2014}.  Experimentally, for the case of superposition of channels, this expansion depends on the physical implementation of the channel. We shall demonstrate that different physical implementations of the same qubit channel can lead to a different expanded description. Notably, the choice of this expanded description of channels completely characterizes the channel behaviour under superposition. This demonstrates that the physical implementation, rather than the act of superposing channels, is the origin of the capacity enhancement.
\begin{figure}
\includegraphics[width=\columnwidth]{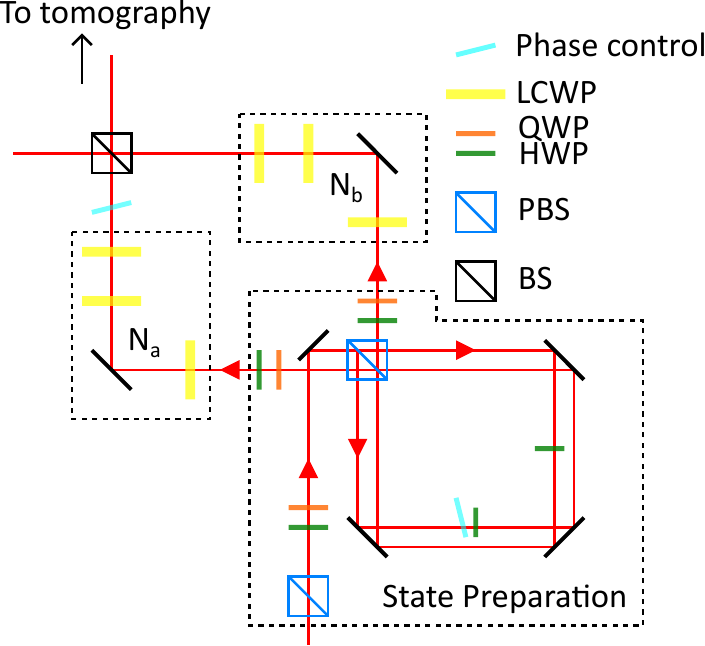}
\caption{The experimental setup is divided into two parts. The photon first goes through the state preparation setup. This setup prepares the photon polarization using a set of half and quarter waveplates (HWP/QWP), which is followed by a Sagnac interferometer that prepares the photon path through a pair of HWPs. The pair of HWPs in the Sagnac is rotated in a correlated manner, and for the preparation of the $\left|+\right>$ path state, the two HWPs are set to be at $\pi/2$ off the horizontal axis. After the Sagnac, the photon goes through a set of HWPs and QWPs to correct for the path-dependent polarization imposed by the Sagnac and restore the photon polarization to the initial preparation. The settings for the sets of HWPs and QWPs immediately after the Sagnac, unlike the initial set of HWPs and QWPs or the HWPs in the Sagnac, is independent of the polarization and path state prepared. The photon then experiences either the channel $N_a$ or $N_b$ depending on the state of the control. Finally, a balanced beam splitter re-interferes the two paths, with state tomography being performed on one of the output ports of the beam splitter. In the order of interaction with an incoming photon, the three LCWPs in channel $N_a$ apply a controlled rotation in the $Z$ axis followed by two rotations in the $X$ axis, whereas the three LCWPs in channel $N_b$ apply a controlled rotation in the $X$ axis followed by two rotations in the $Z$ axis.}
\label{fig:exp_setup}
\end{figure}

This paper is organized in the following manner. We begin by describing our experimental setup in section \ref{sec:setup} followed by experimental results concerning the dependence of the post-selected channel and qutrit channel on the physical implementation of the qubit channel in sections \ref{sec:post_selection} and \ref{sec:qutrit}. Finally, we construct a hierarchy of channel models based on their complexity and completeness in describing the superposed channel in section \ref{sec:hierarchy}, and discuss the implications of our results in section \ref{sec:disc}.

\section{Experimental setup\label{sec:setup}}
The construction of our superposition of channels is based on heralded single photons, with the polarization degree of freedom playing the role of the system qubit and the path degree of freedom playing the role of the control qubit. In our setup, illustrated in figure \ref{fig:exp_setup}, we prepare the photon polarization and path by using a set of waveplates and a Sagnac interferometer respectively. After the preparation, the photon goes into either path \textbf{a} or path \textbf{b} of a Mach-Zehnder interferometer, with the paths corresponding to random unitary channels $N_a$ and $N_b$ respectively. These random unitary channels each consist of three liquid crystal waveplates (LCWP), where the random unitary is implemented by changing the LCWP voltage.  Path \textbf{a} also passes through a glass plate that controls the phase between the two paths, after which the two paths interfere at a beam splitter (BS).

\begin{figure*}[htpb]
    \centering
    \begin{subfigure}[b]{0.3\textwidth}
        \centering
        \caption{}
        \includegraphics[width=\textwidth]{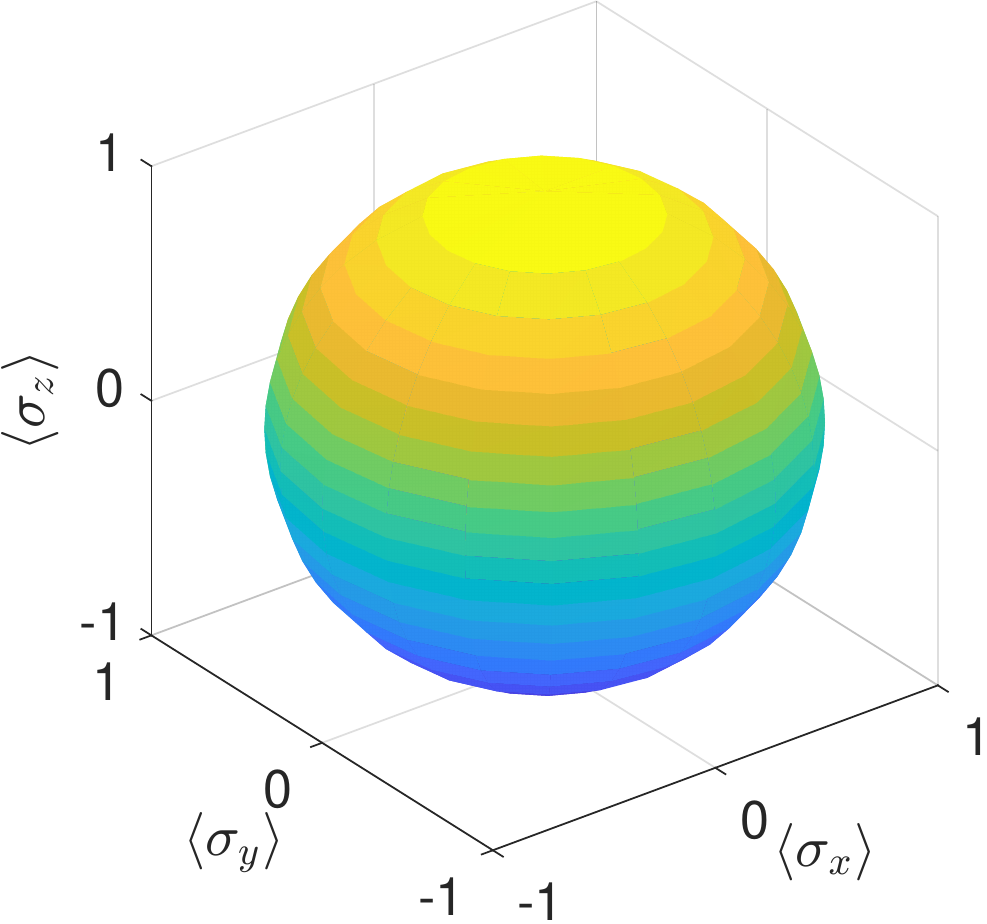}
        \label{fig:post_selected_bloch_Id}
    \end{subfigure}
    \begin{subfigure}[b]{0.3\textwidth}
        \centering
        \caption{}
        \includegraphics[width=\textwidth]{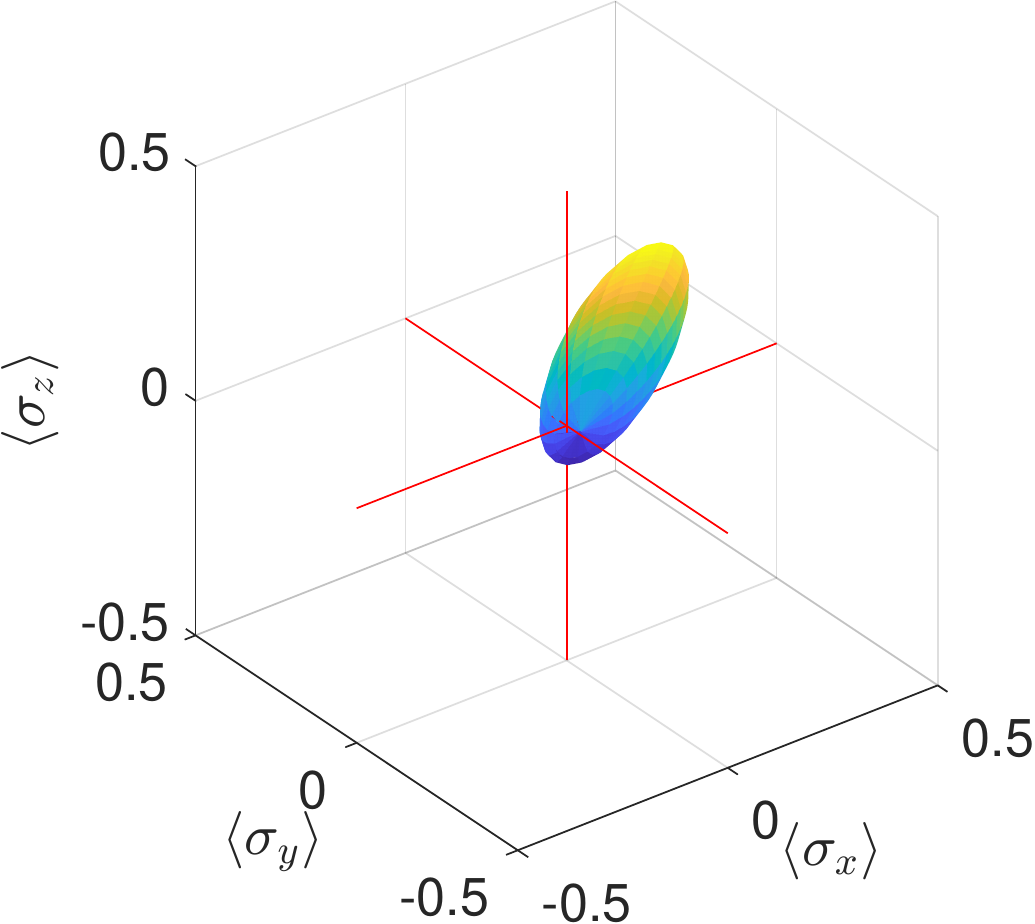}
        \label{fig:post_selected_bloch_Plus}
    \end{subfigure}
    \begin{subfigure}[b]{0.3\textwidth}
        \centering
        \caption{}
        \includegraphics[width=\textwidth]{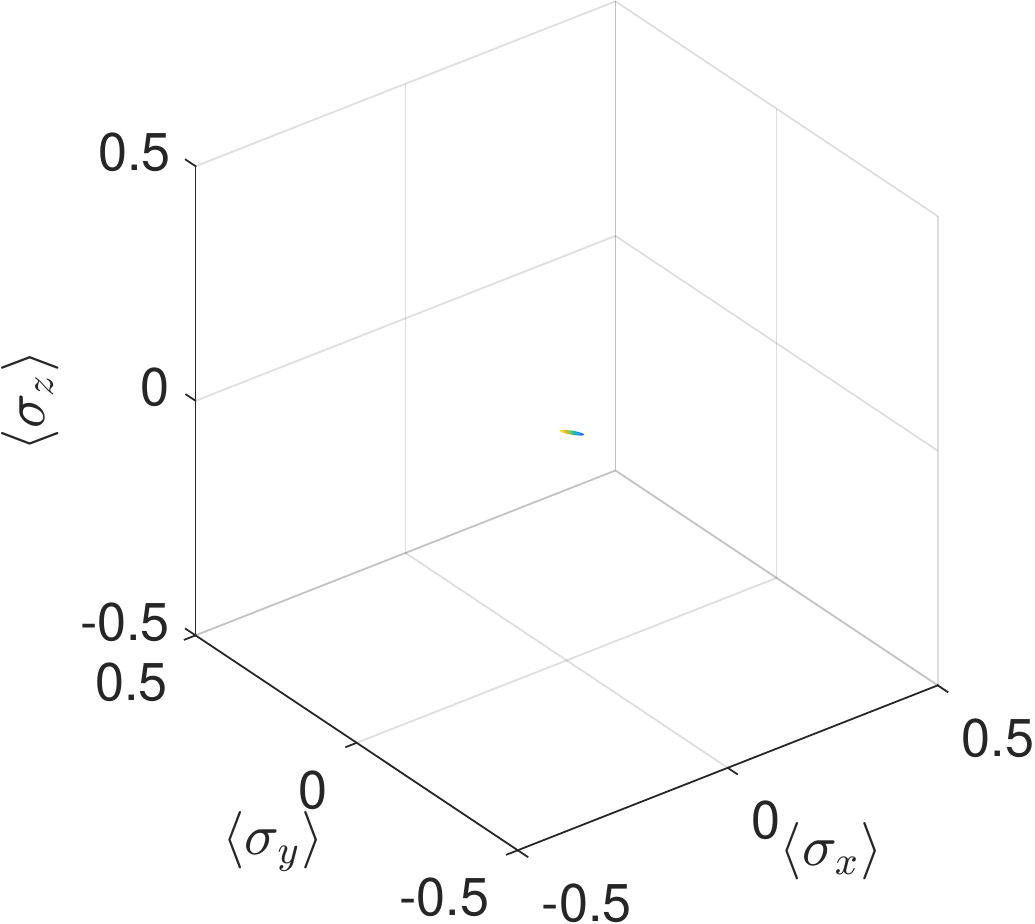}
        \label{fig:post_selected_bloch_Mix}
    \end{subfigure}
    \caption{Bloch sphere representation of the normalized post-selected qubit channel; note the re-scaled axes for the centre and right plots, which run from -0.5 to 0.5. The left plot (\protect\subref{fig:post_selected_bloch_Id}) shows the Bloch sphere representation of an identity channel and can be used as a reference to interpret other Bloch sphere plots. The centre plot (\protect\subref{fig:post_selected_bloch_Plus}) shows the post-selected channel for the phase-coherent implementation with $\alpha=1$, resulting in a Bloch representation that takes the shape of a slanted ellipsoid centred at $(0.10, -0.01, 0.14)$ parameterized by three semi-axes $(0.16, -0.018, 0.15)$, $(0.055, 0.082, -0.017)$, and $(0.011, -0.0088, -0.014)$. Red lines which align with the axis and go through the origin have also been drawn for visual clarity. In theory, this ellipsoid should be an elliptical disc parameterized by the two semi-axes $(\sqrt{2}/9,0,\sqrt{2}/9)$ and $(0,1/9,0)$ centred at $(1/9,0,1/9)$. This deviation from theory is caused by systematics described in appendix \ref{sec:systematics}. The existence of a disc with a definite non-vanishing area is distinct to that of a completely depolarizing channel, such as the one on the right (\protect\subref{fig:post_selected_bloch_Mix}), where the plot shows the post-selected channel for the phase-incoherent implementation, resulting in a point-like shape in its Bloch representation. Theoretically, this plot should represent a completely depolarizing channel and the channel should be shown as a point on the Bloch sphere plot. }
    \label{fig:post_selected_bloch}
\end{figure*}

\begin{figure*}[htpb]
    \centering
    \includegraphics[width=\columnwidth]{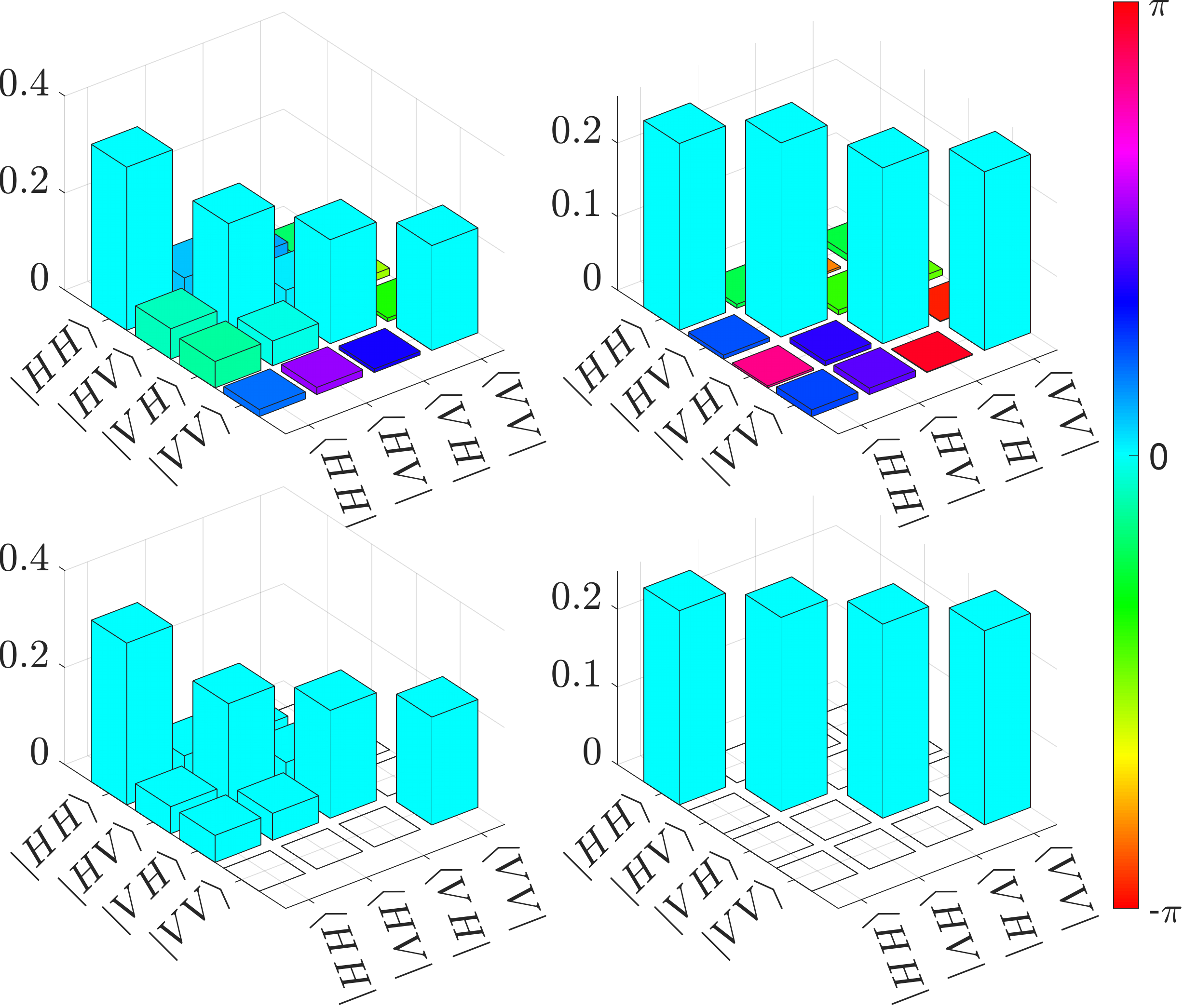}
    \caption{Plot of the density matrix for normalized post-selected qubit channel dual state with $N_a$ and $N_b$ for the phase-coherent (incoherent) implementation with $\alpha=1$ on the left (right) with experimental results on the top and theory at the bottom. The dual state Hilbert space is labelled by two labels $H$ or $V$ first denoting the input polarization followed by the output polarization. The height of the bars and their colours represent the amplitudes and phase respectively of each element. Without phase mixing (left plots), post-selection partially restores coherence to the channel map. It also preferentially transmits H polarization, as indicated by the larger amplitude in $\left|HH\right\rangle\left\langle HH\right|$ than the rest of the diagonal elements. With phase mixing (right plots), post-selection does not restore any coherence, with the resulting channel being a completely depolarizing channel.}
    \label{fig:post_selected_map}
\end{figure*}

The unitary performed by the Mach-Zehnder is given by
\begin{equation}\label{eq:two_qubit_op}
    \hat{U}_{i,j}^{(MZ)}=\left(\left|\textbf{a}\right>\left<\textbf{a}\right|\otimes \hat U_i^{(a)} + \left|\textbf{b}\right>\left<\textbf{b}\right|\otimes \hat U_j^{(b)}\right)
\end{equation}
where $\hat U_i^{(a)}$ and $\hat U_j^{(b)}$ are the unitary operators corresponding to the polarization rotation given by LCWPs in $N_a$ and $N_b$,  $\left|\textbf{a}\right>\left<\textbf{a}\right|$ and $\left|\textbf{b}\right>\left<\textbf{b}\right|$ corresponds to the projectors on the path qubit, and the indices $i,j$ denote elements in the set of all possible unitary operators. To simulate the effects of a random unitary channel, we perform our experiment using all possible combinations of unitaries $\{U_i^{(a)}\}_i$ and $\{U_j^{(b)}\}_j$ and taking a weighted average of the results with weightings $p_i^{(a)}$ and $p_j^{(b)}$ that correspond to the probabilities of those unitaries in the respective random unitary channel. This effectively emulates random unitary channels for $N_a$ and $N_b$, with Kraus operators given by $\hat{K}_i^{(a)}=\sqrt{p_i^{(a)}}\hat U_i^{(a)}$ and $\hat{K}_j^{(b)}=\sqrt{p_j^{(b)}}\hat U_j^{(b)}$. The overall channel given by our Mach-Zehnder interferometer acting on some input path and polarization state $\rho^{(MZ)}$ is then also a random unitary channel, given by
\begin{multline}
    \boldsymbol{\Phi}^{(MZ)}\left(\rho^{(MZ)}\right)\\
    =\sum_{i,j} p_i^{(a)}p_j^{(b)}\hat{U}_{i,j}^{(MZ)} \rho^{(MZ)} \hat{U}_{i,j}^{(MZ)\dagger},
\end{multline}
with the summation operation summing over the entire set of possible unitary operators $\hat U_i^{(a)}$ and $\hat U_j^{(b)}$ weighted over the probability of those unitaries $p_i^{(a)}p_j^{(b)}$.

To understand the nature of the implementation-dependence of superposing qubit channels, we will compare the superposed channels created by superposing two different implementations of the depolarizing channel.
First, we remind readers of some mathematical properties of the depolarizing channels.
We note that a set of Kraus operators in the qubit Hilbert space ${\hat{K}_i}$ can be expressed in terms of the qubit operator basis formed by the identity and the three Pauli operators, such that
\begin{equation}
    \sum_{k,l}d_{k,l}\,\hat{\sigma}_k \rho \hat{\sigma}_l^\dagger,
\end{equation}
\begin{equation}\label{eq:decomp}
    d_{k,l} = \sum_i Tr\left[\hat{K}_i\hat{\sigma}_k^\dagger\right]\cdot  Tr\left[\hat{K}_i\hat{\sigma}_l^\dagger\right]^*/4,
\end{equation}
where $\hat{\sigma}_1$, $\hat{\sigma}_2$, and $\hat{\sigma}_3$ are Pauli matrices (and will be used interchangeably with $\hat{\sigma}_x$, $\hat{\sigma}_y$, and $\hat{\sigma}_z$), and $\hat{\sigma}_0$ is the identity.
The channel formed by the set of Kraus operators $\{\hat{K}_i\}_{i}$ describes a depolarizing channel if it can be described by a random unitary channel, where $\hat{\sigma}_1$, $\hat{\sigma}_2$, and $\hat{\sigma}_3$ are performed with probability $\alpha/4$ and the identity operator with probability $1 - 3\alpha/4$. Mathematically, the decomposition of a depolarizing channel from equation \ref{eq:decomp} gives
\begin{equation}
    d_{k,l} =  
    \begin{cases}
        1 - 3\alpha/4 &k=l=0\\
        \alpha/4 &k=l\neq 0\\
        0 &k\neq l
    \end{cases}
\end{equation}
where $\alpha$ is the degree of depolarization, with $\alpha=1$ corresponding to a completely depolarizing channel.

We will denote the two implementations of the depolarizing channel as the phase-coherent and phase-incoherent implementations, with all implementation-specific symbols being denoted with the unbracketed superscript $coh$ and $inc$ respectively. 
More precisely, in the phase-coherent implementation, four different operators are randomly implemented for each channel. These operators are, for $N_a$ and $N_b$ respectively,

\begin{align}
&\begin{aligned}\label{eq:imp1}
    \{\hat{K}_i^{(a),coh}\}_{i}=\{
    \sqrt{p_0}s_{0}^{(a)}\hat {\sigma}_0,\sqrt{p_1}s_{1}^{(a)}\hat {\sigma}_1,\\
    \sqrt{p_2}s_{2}^{(a)}\hat {\sigma}_2,\sqrt{p_3}s_{3}^{(a)}\hat {\sigma}_3\};\\
    \{\hat{K}_j^{(b),coh}\}_{j}=\{
    \sqrt{p_0}s_{0}^{(b)}\hat {\sigma}_0,\sqrt{p_1}s_{1}^{(b)}\hat {\sigma}_1,\\
    \sqrt{p_2}s_{2}^{(b)}\hat {\sigma}_2,\sqrt{p_3}s_{3}^{(b)}\hat {\sigma}_3\}.\\
\end{aligned}
\end{align}
where $\hat U_i^{(a)}=s_{i}^{(a)}\hat {\sigma}_i$ and $\hat U_j^{(b)}=s_{j}^{(b)}\hat {\sigma}_j$, and $s_i^{(a)}$ and $s_j^{(b)}$ are the phase factors associated with each Pauli operators in the phase-coherent implementation. The probabilities $p_m$ and phase factors $s_i^{(a)}$ and $s_j^{(b)}$ are given by
\begin{equation}\label{eq:pauliProb}
    p_m =  
    \begin{cases}
        1 - 3\alpha/4 &m=0;\\
        \alpha/4 &m\neq 0,
    \end{cases}
\end{equation}
\begin{align}
&\begin{aligned}
    s_{m}^{(a)} &= s_{m}^{(b)} =  1 \text{ for }\,m\neq 2;\\
    s_{2}^{(a)} &=i;\\
    s_{2}^{(b)} &=-i.
\end{aligned}
\end{align}
The $s_{2}^{(a)}$ and $s_{2}^{(b)}$ phase factors correspond to the configuration of the LCWPs in $N_a$ and $N_b$ in our physical setup, where a $Z$ aligned LCWP comes before the $X$ aligned LCWP in $N_a$ and the reverse is true for $N_b$. 

In the phase-incoherent implementation, we completely randomize the path phase by add four additional operators to the two operator sets in the phase-coherent implementation. These four additional operators correspond to the operators in the phase-coherent implementation with an additional $\pi$ phase. The phase-incoherent implementation, therefore, has operators
\begin{align}
&\begin{aligned}\label{eq:imp2}
    \{\hat{K}_i^{(a),inc}\}_{i}&=\{\hat{K}_j^{(b),inc}\}_{j}=\\
    \{&\sqrt{(p_0/2)}\,s_{0}^{(a)}\hat {\sigma}_0,\sqrt{(p_1/2)}\,s_{1}^{(a)}\hat {\sigma}_1,\\
    &\sqrt{(p_2/2)}\,s_{2}^{(a)}\hat {\sigma}_2,\sqrt{(p_3/2)}\,s_{3}^{(a)}\hat {\sigma}_3,\\
    -&\sqrt({p_0/2)}\,s_{0}^{(a)}\hat {\sigma}_0,-\sqrt{(p_1/2)}\,s_{1}^{(a)}\hat {\sigma}_1,\\
    -&\sqrt{(p_2/2)}\,s_{2}^{(a)}\hat {\sigma}_2,-\sqrt{(p_3/2)}\,s_{3}^{(a)}\hat {\sigma}_3\}.
\end{aligned}
\end{align}
Note in the phase-incoherent implementation, the random unitary operators implemented in $N_a$ and $N_b$ are not necessarily equal, but rather are drawn independently from the same set of unitaries.  Experimentally, this $\pi$ phase is implemented by the third LCWP with an optical axis aligned perpendicularly to the second LCWP. By tuning the voltages of both LCWP simultaneously, a polarization-independent phase shift can be applied to the photons. This phase acts as a global phase when channels $N_a$ and $N_b$ are not in superposition.

\section{Implementation Dependence\label{sec:post_selection}}
A distinct feature of superposing channels is the dependence of the superposed channel map on the operators used to implement the non-superposed channels used in the superposition. As a consequence, the choice of using implementations \ref{eq:imp1} or \ref{eq:imp2} for the simple qubit channels $N_a$ and $N_b$ will result in two different versions of the two-qubit channel $\boldsymbol{\Phi}^{(MZ)}\left(\rho^{(MZ)} \right)$. 

When the path qubit is prepared and post-selected on $\left|+\right>=\frac{1}{\sqrt{2}}\left(\left|\textbf{a}\right>+\left|\textbf{b}\right>\right)$, the resulting post-selected qubit channel
\begin{equation}
    \boldsymbol{\Phi}^{(p)}\left(\rho^{(p)}\right)=\sum_{i,j} \hat{K}_{i,j}^{(p)} \rho^{(p)} \hat{K}_{i,j}^{(p)\dagger},
\end{equation}
will have Kraus operators $\hat{K}_{i,j}^{(p)}$ given by 
\begin{equation}\label{eq:post_selected_op}
    \hat{K}_{i,j}^{(p)}=\sqrt{p_i^{(a)}p_j^{(b)}}\,\frac{\hat U_i^{(a)} + \hat U_j^{(b)}}{2}.
\end{equation}
The implementation-dependence of the post-selected qubit channel can be seen as a consequence of $\hat{K}_{ij}^{(p)}$ being sensitive to  the global phases, which are reflected in $s_{i}^{(a)}$ and $s_{j}^{(b)}$, of the two channels $N_a$ and $N_b$. These global phases of the two channels become phases between states $\left|\textbf{a}\right>$ and $\left|\textbf{b}\right>$ in the control qubit when the channels are put in a superposition.
The Pauli decomposition of the post-selected channel helps us illuminate the nature of the post-selected channel's dependence on channel implementation. For the phase-coherent implementation, the Pauli decomposition of the set of Kraus operators $\hat{K}_{i,j}^{(p)}$ given by $f_{k,l} = \sum_{i,j} Tr\left[\hat{K}_{i,j}^{(p)}\hat{\sigma}_k^\dagger\right]\cdot  Tr\left[\hat{K}_{i,j}^{(p)}\hat{\sigma}_l^\dagger\right]^*/4$ results in 
\begin{equation}\label{eq:imp1paulidecomp}
    f_{k,l}^{coh} =  
    \begin{cases}
        \frac{p_k}{2}+\frac{p_k^2}{2}\,\text{Re}\left[s_{k}^{(a)}s_{k}^{(b)*}\right] &k=l;\\
        \frac{p_k p_l}{4}\left(s_{k}^{(a)}s_{l}^{(b)*}+s_{k}^{(b)}s_{l}^{(a)*}\right) &k\neq l.
    \end{cases}
\end{equation}
The phase-dependence of the post-selected channel can thus be seen by noting equation \ref{eq:imp1paulidecomp}'s dependence on the $s_{k}^{(a)}s_{l}^{(b)*}$ terms.

Examining figure \ref{fig:post_selected_bloch}, which shows the Bloch sphere representation of the post-selected channel, we can see that the eigenstate with the positive eigenvalue for the Hadamard unitary $\hat H=1/\sqrt{2}\left(\hat \sigma_x + \hat \sigma_z\right)$ experiences the least amount of depolarization from the post-selected channel in the phase-coherent implementation. An explanation for this is given in appendix \ref{sec:ps_noise_explained}.

In contrast, the phase-incoherent implementation has a Pauli decomposition that yields
\begin{equation}\label{eq:imp2paulidecomp}
    f_{k,l}^{inc} =  
    \begin{cases}
        \frac{p_k}{2} &k=l;\\
        0 &k\neq l.
    \end{cases}
\end{equation}
The inclusion of a duplicate set of operators with a $\pi$ phase shift causes the contribution of the operator phase components $s_{k}^{(a)}$ and $s_{l}^{(b)*}$ to cancel out in the Pauli decomposition. Substituting the probabilities of equation \ref{eq:pauliProb} into \ref{eq:imp2paulidecomp}, we have $f_{k,l}^{inc}=\frac{1}{2}d_{k,l}$. This indicates that the post-selected channel given by the phase-incoherent implementation results in a depolarizing channel with a post-selection probability of $1/2$, exactly the same as the channels $N_a$ and $N_b$.

Figures \ref{fig:post_selected_bloch} and \ref{fig:post_selected_map} highlight this implementation-dependence quantitatively through the tomographic reconstruction of the phase-coherent and phase-incoherent channels, and qualitatively in the channels' shape on the Bloch sphere. As seen from the figures, for the phase-coherent implementation with Kraus operators $\{\hat{K}_i^{(a),coh}\}_{i}$ and $\{\hat{K}_i^{(b),coh}\}_{j}$ with $\alpha=1$, the post-selection results in a channel that is not completely depolarizing, even though the individual channels themselves are. This is in contrast to the phase-incoherent implementation with Kraus operators $\{\hat{K}_i^{(a),inc}\}_{i}$ and $\{\hat{K}_j^{(b),inc}\}_{j}$, where due to the randomization of channel phases, the effect of post-selection reduces to a classical mixture of depolarizing channels, resulting in a completely depolarizing channel. The result of superposing two phase-incoherent implementations also extends to the superposition of one phase-coherent and one phase-incoherent implementation.

\begin{figure*}[htpb]
    \centering
    \includegraphics[width=0.7\textwidth]{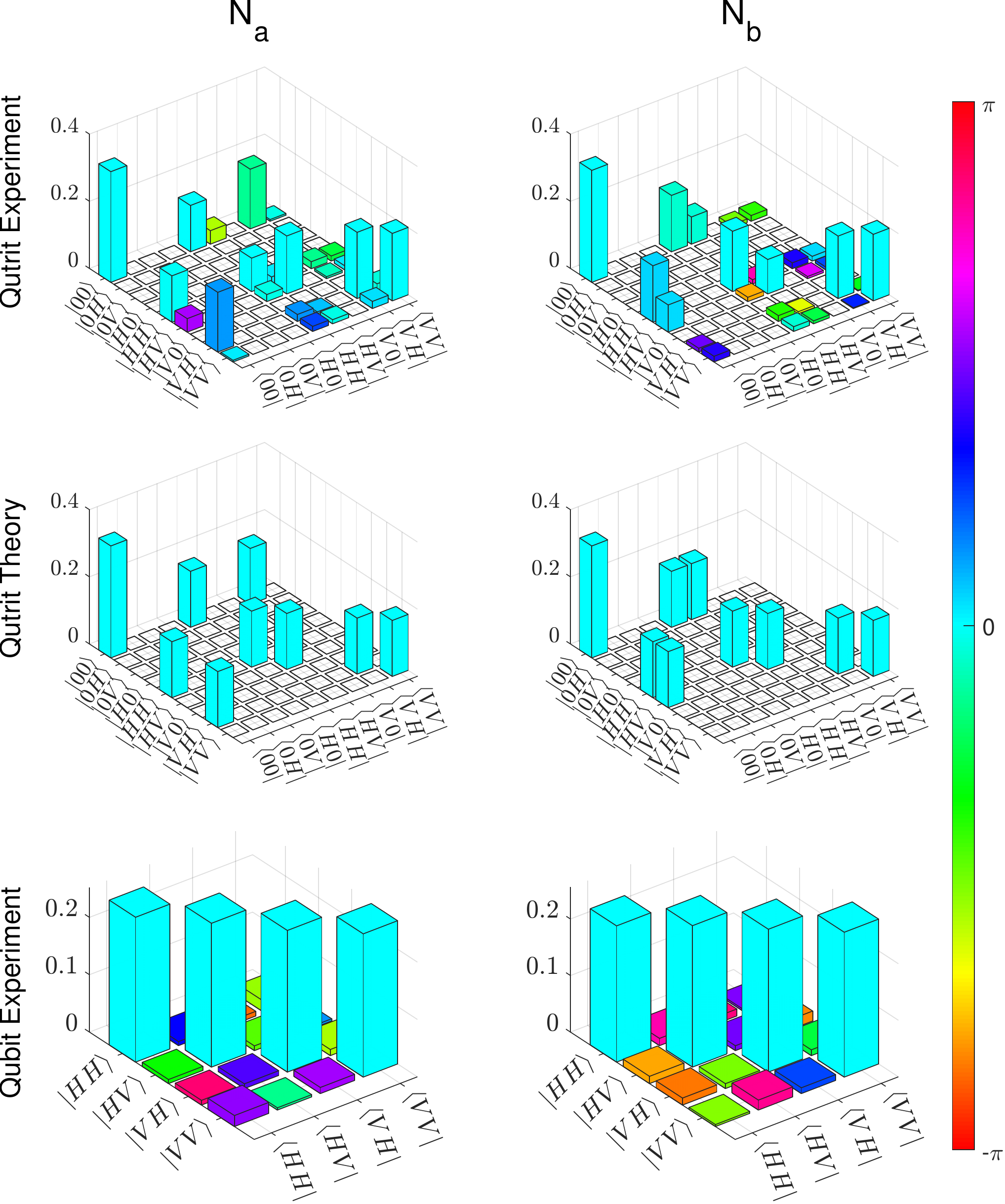}
    \caption{Plot of channel dual state for $N_a$ ($N_b$) for the phase-coherent implementation with the degree of depolarization $\alpha=1$ on the left (right) with the qutrit model in the top row, the qutrit model from theory in the middle row and qubit model in the bottom row. The vacuum state is labelled as zero for compactness. The difference between the channels $N_a$ and $N_b$ is highlighted in the qutrit models by the coherence between $\left|00\right\rangle$ and $\left|VH\right\rangle$ ($\left|HV\right\rangle$), despite them both having an identical qubit model description.  It is important to note that any terms associated with an input polarization state and an output zero state or vise-versa (e.g. $\left|0V\right>$ or $\left|H0\right>$) is set to zero in our fitting model. Given our method of tomography, an input polarization state becoming an output zero (or vise-versa) corresponds to a photon disappearing in one path of the Mach Zehnder and reappearing in the other path, which does not correspond to any physical mechanism present in our experiment. Appendix \ref{sec:systematics} elaborates on the systematic errors that contribute to the deviations between experiment and theory. }
    \label{fig:qutrit_map}
\end{figure*}

\begin{figure*}[htpb]
    \centering
    \includegraphics[width=0.7\textwidth]{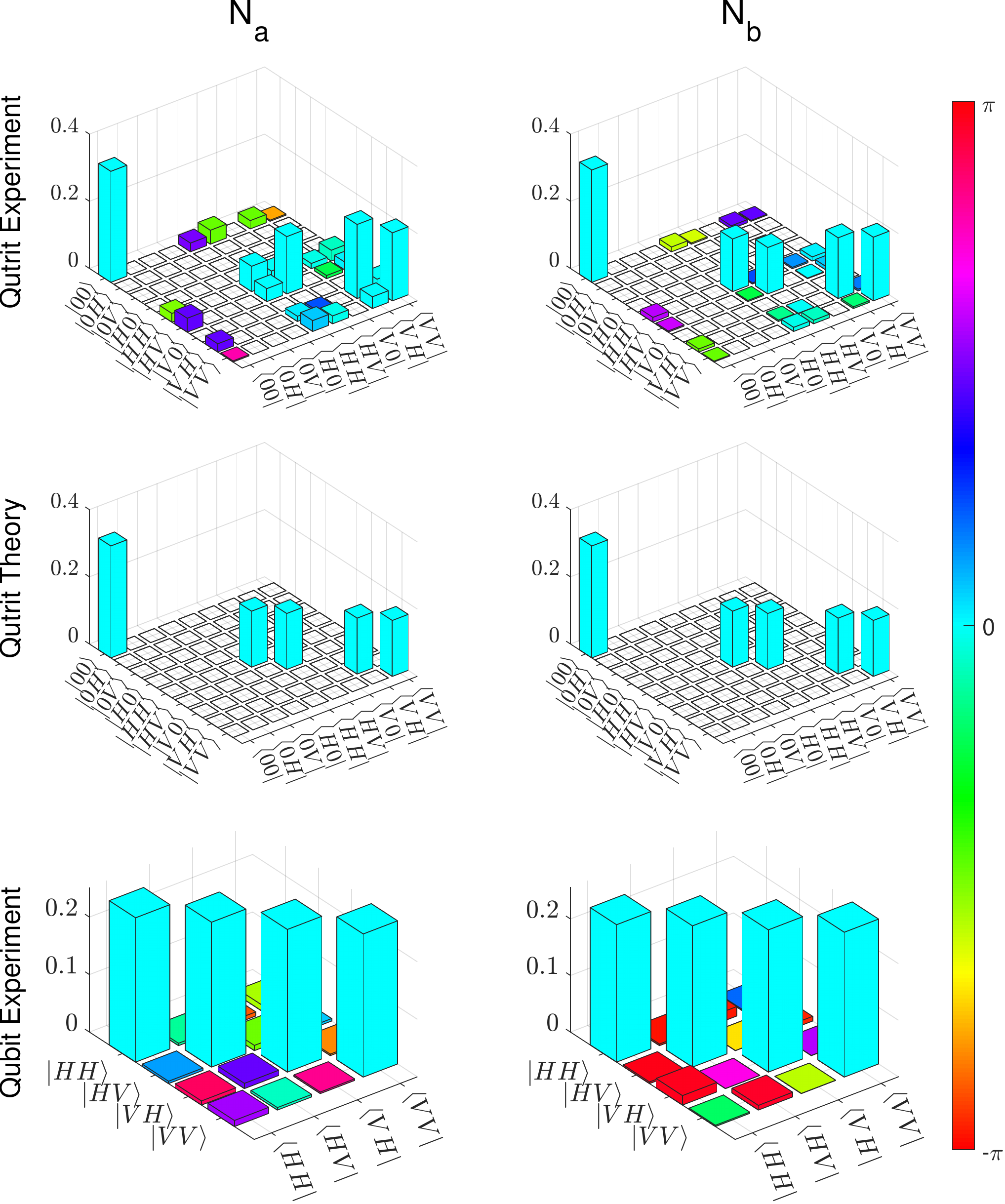}
    \caption{Plot of qutrit channel dual state with $N_a$ ($N_b$) for the phase-incoherent implementation with the degree of depolarization $\alpha=1$ on the left (right)  with the qutrit model in the top row, the qutrit model from theory in the middle row and qubit model in the bottom row. The phase randomization of the qubit channels manifests as a loss of coherence between $\left|00\right\rangle$ and $\left|VH\right\rangle$ ($\left|HV\right\rangle$) in the qutrit map, resulting in a purely diagonal channel dual state. Appendix \ref{sec:systematics} elaborates on the systematic errors that contribute to the deviations between experiment and theory.}
    \label{fig:qutrit_mixing_map}
\end{figure*}

\section{Qutrit Model\label{sec:qutrit}}
\begin{figure}
    \centering
    \includegraphics{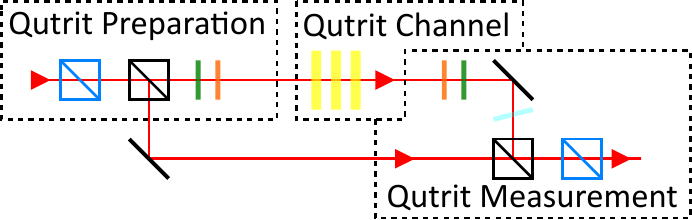}
    \caption{The conceptually equivalent setup for qutrit channel tomography. This setup allows the preparation and measurement of a quantum state spanned by the basis state $|0\rangle$ (which is equivalent to having the photon in the interferometer path without the qutrit channel), $|V\rangle$, and $|H\rangle$. In the qutrit preparation stage, $|H\rangle$ polarized light is sent to a beamsplitter, with one arm of the beamsplitter sent to an HWP and a QWP which turns the $|H\rangle$ polarized light into arbitrary polarizations. A similar but reversed setup is used for qutrit measurement. Phase between $|0\rangle$ and the polarization states for both the preparation and measurement is tuned by a common phase plate, which is possible due to the assumption that the qutrit channel does not induce amplitude exchange between the $|0\rangle$ state and the polarization states. Finally, a tomographically complete preparation and measurement can be accomplished by blocking one or none of the paths in the interferometer. }
    \label{fig:qutrit_conceptual}
\end{figure}

The inability of the simple qubit channel map to translate directly to a unique description of the superposed channel demonstrates that it is insufficient for describing the physical operations of the channel under superposition. This is because of the introduction of the control qubit, which requires that we can effectively `turn off' a quantum channel by sending in zero photons. Thus, apart from the description of the channel's action on the polarization qubit, an additional description of the channel action on the vacuum state needs to be included, as noted by \cite{Rubino2021,Abbott2020,Guerin2019}, which in turn results in the simple qubit channels becoming qutrit channels: two states from the original polarization qubit channel, and one new state corresponding to the vacuum (`off') state. 
Here, we perform qutrit channel tomography for the aforementioned qutrit channels explicitly through the procedures described in appendix \ref{sec:qutrit_tomo}. Our qutrit Hilbert space consists of the vacuum (zero photon) state $|0\rangle$, the polarization state H $|H\rangle$, and the polarization state V $|V\rangle$. When performing tomography for channel $N_a$($N_b$), the preparation for the $|0\rangle$ state is equivalent to not sending the photon into the channel $N_a$($N_b$). Conceptually, the qutrit tomography procedures described in appendix \ref{sec:qutrit_tomo} are equivalent as follows. The preparation of the qutrit state is done by first preparing the photon in the H polarization and path state in either $|\textbf{a}\rangle$ or $|\textbf{b}\rangle$. Then, for the channel tomography for $N_a$($N_b$), we vary the photon polarization only if the photon is in the $|\textbf{a}\rangle$($|\textbf{b}\rangle$) path state. This has the effect of preparing a state in the Hilbert space spanned by $|0\rangle$, $|V\rangle$, and $|H\rangle$. For the tomography of channel $N_a$($N_b$), channel $N_b$($N_a$) is always set to the identity channel. Similar to the input qutrit preparation, the qutrit is measured by performing polarization measurement only on the $|\textbf{a}\rangle$($|\textbf{b}\rangle$) path and a measurement of the path qubit. A setup equivalent to the qutrit tomographic procedure is outlined in figure \ref{fig:qutrit_conceptual}.

Figures \ref{fig:qutrit_map} and \ref{fig:qutrit_mixing_map} show the tomographic results for qutrit channels of $N_a$ and $N_b$ for both implementations with $\alpha=1$. In these figures, the effects of the operator phase $s_i$ of the two implementations can be observed in the coherences between the polarizations and the vacuum state. Both channels in the phase-coherent implementation partially preserve coherence between $\left|00\right\rangle$ and $\left|HH\right\rangle$ but differ by the coherence between $\left|00\right\rangle$ and $\left|VH\right\rangle$ (which is partially preserved for $N_a$ and not for $N_b$), and the coherence between $\left|00\right\rangle$ and $\left|HV\right\rangle$ (which is partially preserved for $N_b$ and not for $N_a$). On the other hand, neither channel preserves any coherences in the phase-incoherent implementation, resulting in a purely diagonal channel dual state.

\section{Channel Hierarchy\label{sec:hierarchy}}
Fundamentally, the simple qubit ($\boldsymbol{\Phi}^{(a)}\left(\rho^{(a)}\right)$ and $\boldsymbol{\Phi}^{(b)}\left(\rho^{(b)}\right)$), post-selected ($\boldsymbol{\Phi}^{(p)}\left(\rho^{(p)}\right)$), Mach-Zehnder ($\boldsymbol{\Phi}^{(MZ)}\left(\rho^{(MZ)}\right)$), and qutrit ($\boldsymbol{\Phi}^{(T0)}\left(\rho^{(T0)}\right)$ and $\boldsymbol{\Phi}^{(T1)}\left(\rho^{(T1)}\right)$) channel model(s) are descriptions of the same setup with differing levels of detail. It is therefore helpful to order those different descriptions based on these levels. The simple qubit and post-selected channels can be derived from the Mach-Zehnder channel by preparing and measuring the control in the appropriate state, and the Mach-Zehnder channel can in turn be derived by preparing and measuring the input and output in the one-photon subspace of the two-qutrit channel $\boldsymbol{\Phi}^{(2T)}$ given by
\begin{equation}
    \boldsymbol{\Phi}^{(2T)}=\boldsymbol{\Phi}^{(T0)}\otimes\boldsymbol{\Phi}^{(T1)}.
\end{equation}
We do not have direct experimental access to the two-qutrit channel as the design of our setup does not allow us to access $\boldsymbol{\Phi}^{(T0)}$ and $\boldsymbol{\Phi}^{(T1)}$ simultaneously. Nonetheless, it is important to discuss it conceptually in this section. Based on the channel description's level of detail, the channel models can be partially ordered in the following way
\begin{subequations}
\begin{eqnarray}
    \boldsymbol{\Phi}^{(2T)}\succ\boldsymbol{\Phi}^{(MZ)}\succ\boldsymbol{\Phi}^{(a)},\boldsymbol{\Phi}^{(b)},\\
    \boldsymbol{\Phi}^{(2T)}\succ\boldsymbol{\Phi}^{(T0)}\succ\boldsymbol{\Phi}^{(a)},\\
    \boldsymbol{\Phi}^{(2T)}\succ\boldsymbol{\Phi}^{(T1)}\succ\boldsymbol{\Phi}^{(b)},
\end{eqnarray}
\end{subequations}
where any channel description ordered lower on the partial ordering can be extracted from a channel description ordered higher on the partial ordering, either by taking a subspace or post-selecting on the channel higher on the ordering. More broadly speaking, our partial ordering here is a specific kind of channel divisibility \cite{Heinosaari2017,Duarte2022} that involves dimensional reduction.
\begin{figure*}[htpb]
    \centering
    \includegraphics[width=1\textwidth]{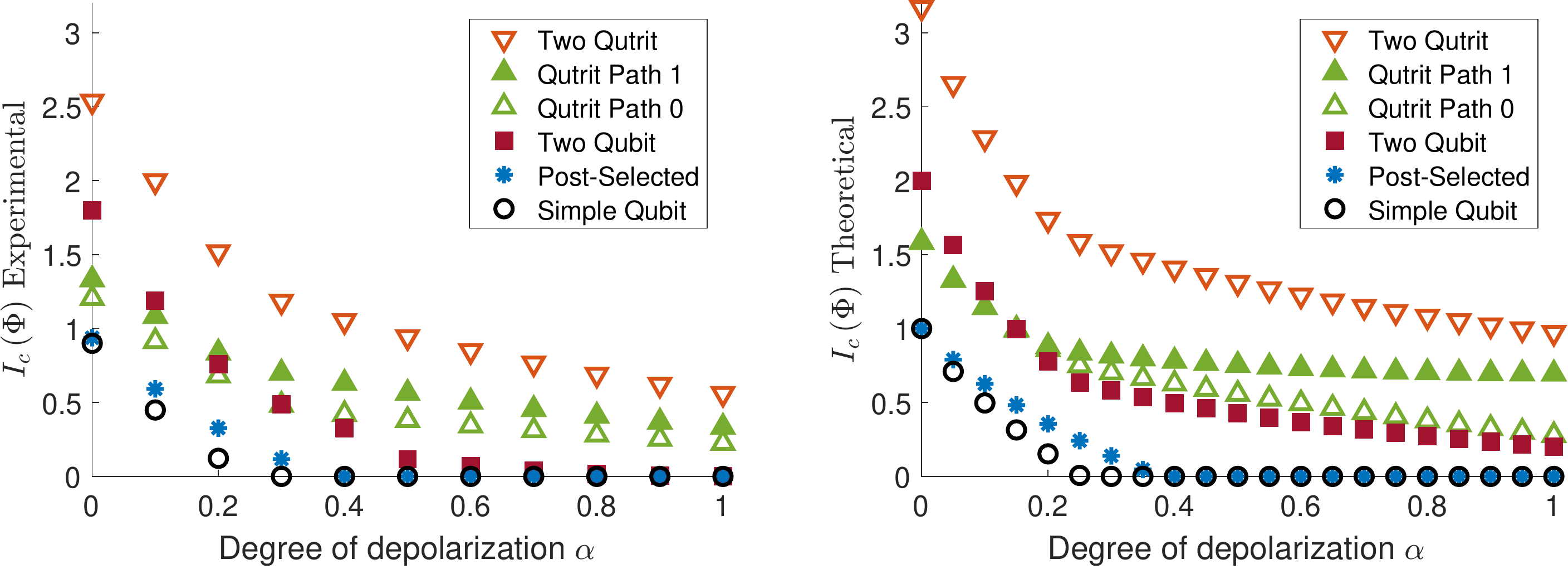}
    \caption{The maximum coherent information for different channel models in the phase-coherent implementation. The plot for the two-qutrit channel in both theory and experiment is calculated by adding the maximum coherent information for qutrit path 0 and path 1 instead of optimizing over possible input state as we did for the other channels. The two qutrit channel coherent information should therefore be interpreted as the lower bound for the maximum coherent information. Nonetheless, the plot still indicates that the channels follow the established hierarchy in that no channel model higher in the partial ordering has less maximum coherent information than a channel model lower in the ordering. This is true for both experiment and theory.}
    \label{fig:Coh_info}
\end{figure*}

With this partial ordering in mind, we compare the maximum coherent information of our depolarizing channels under different channel models and implementations. The maximum coherent information is related to the ability of a quantum channel to generate entanglement across two parties and can be understood as the amount of quantum information that can be transmitted through a single use of the channel \cite{watrous_2018}. We have chosen to use this quantity over classical channel capacity and quantum channel capacity due to the former not capturing the quantum characteristics of channels and the latter being hard to compute. The maximum coherent information is given by
\begin{multline}
    I_c\left(\boldsymbol{\Phi}\right)=\\\max_{\{\rho\}}[H\left(\boldsymbol{\Phi}\left(\rho\right)\right)\\
    -H\left(\boldsymbol{\left(\Phi\otimes\mathbb{I}\right)}\left(vec\left(\sqrt\rho\right)\cdot vec\left(\sqrt\rho\right)^\dagger\right)\right)],
\end{multline}
where $vec\left(\sqrt\rho\right)$ is the purification of the state $\rho$ and $H$ is the von Neumann entropy, given by
\begin{equation}
    H\left(\rho\right)=\text{Tr}\left[\rho\,\text{log}\left(\rho\right)\right].
\end{equation}
 Figure \ref{fig:Coh_info} shows the maximum coherent information at various levels of depolarization. The capacity enhancement when simple qubit channels are post-selected in a superposition can be seen at $\alpha$ going from 0 to approximately 0.4, where the maximum coherent information from the post-selected channel is strictly greater than that of the qubit channel.

It is important to stress that the maximum coherent information for the two-qutrit channel, unsurprisingly, is strictly greater than all other channel models under investigation. Therefore, no capacity enhancement would be found if one used the appropriate two-qutrit channel model or superposition channel model to describe the post-selected channel.

\section{Discussion\label{sec:disc}}
The capacity enhancement in coherently controlled channels may seem surprising at first, yet this surprise is perhaps due to the implicit assumption that a coherently controllable channel can be trivially implemented using its non-controlled counterpart. This is not true, as the inclusion of the control qubit necessitates an expanded description of the channels to accommodate the transmission of the extra control qubit. For the case of superposing qubit channels, the channel action on vacuum needs to be accounted for. To illustrate this, we have experimentally reconstructed relevant qutrit channels for three different implementations of the depolarizing channel. Indeed, for depolarizing channels, the quantum information that one can send through the superposition is strictly less than that of the relevant qutrit channels. It is therefore more appropriate to attribute the increase in channel capacity of superposing channels to the required expansion of the channel's input and output Hilbert space, rather than the act of superposition itself.

While the implementation-dependence of coherently controlled channels on its non-coherently controlled counterpart is a feature of the superposed channel that is not shared for all coherently controlled channels, all coherent control schemes require the transmission of an extra control qubit. Our work provides insight into how channel expansion contributes to the capacity enhancement in coherently controlled channels.

\begin{acknowledgments}
This work was supported by grant number ``FQXi FFF Grant number FQXi-RFP-1819'' from the Foundational Questions Institute and Fetzer Franklin Fund, a donor advised fund of Silicon Valley Community Foundation; the Natural Sciences and Engineering Research Council (NSERC) of Canada; and CIFAR. It also made use of some equipment purchased with the assistance of the Fetzer Franklin Fund of the
John E. Fetzer Memorial Trust. A.M.S is a fellow of CIFAR.
\end{acknowledgments}

\appendix
\section{LCWP Characterization}
We used a total of six LCWPs in our experiment, two from Thorlabs and four from Meadowlark Optics. They are distributed in Mach-Zehnder interferometer such that there are two Meadowlark LCWPs and one Thorlabs LCWP in each arm, and ordered such that incoming light always goes through the two Meadowlark LCWPs first. The voltage to phase retardance for each LCWPs is calibrated by sending horizontally polarized light through the LCWP with its optical axis at 45 degrees, with measurement of the subsequent light in the horizontal/vertical basis. Input voltage-dependent absorption was also characterized, and was determined to be at most a $3\%$ change.
\section{Photon Souce}
We generated $808 nm$ SPDC photons using PPKTP type-II co-linear down-conversion, where one of the photons is used as a herald, resulting in $\sim 63,000$ heralded single photons per second. After the single-mode fibre-coupling into a detector, we detect $\sim 8,000$ heralded single photons per second, with losses due to unwanted absorption, reflection, and single-mode coupling inefficiencies.
\section{Systematic errors}\label{sec:systematics}
Deviations between the theoretical and experimental results in figures \ref{fig:post_selected_bloch}, \ref{fig:qutrit_map} and \ref{fig:qutrit_mixing_map} are dominated by three sources of systematic error. The first is due to phase fluctuation in our Mach-Zehnder interferometer of about $0.3$ $rad$ (comparable to, in figure \ref{fig:qutrit_map} the phase error between $\left|VH\right\rangle$ and $\left\langle00\right|$ of $~0.43$ $rad$ for the phase-coherent implementation). The second is the calibration error in the polarization rotation axis of the LCWPs of about $0.06$ $rad$. Finally, there is a polarization dependence absorbance of at most $\sim 7\%$ present in our setup.
\section{Post-selected channel map of phase-coherent implementation} \label{sec:ps_noise_explained}
\begin{table*}[]
\centering
\begin{tabular}{cc|cccc|}
\cline{3-6}
                                                                                              &                     & \multicolumn{4}{c|}{$\hat{U}_i$}                                                                                                                                                                                                                         \\ \cline{3-6} 
                                                                                              &                     & \multicolumn{1}{c|}{$\hat {\sigma}_0$}                   & \multicolumn{1}{c|}{$\hat {\sigma}_1$}                             & \multicolumn{1}{c|}{$\hat {\sigma}_3$}                             & $i\hat {\sigma}_2$                            \\\hline
\multicolumn{1}{|c|}{\multirow{4}{*}{\rotatebox[origin=c]{90}{$\hat{U}_j$}}} & $\hat {\sigma}_0$   & \multicolumn{1}{c|}{$2 \hat {\sigma}_0$}                 & \multicolumn{1}{c|}{$\hat {\Pi}_{\left|+\right>}$}                 & \multicolumn{1}{c|}{$\hat {\Pi}_{\left|0\right>}$}                 & $\sqrt{2} \hat {H} \cdot \hat {\sigma}_1$           \\ \cline{2-6} 
\multicolumn{1}{|c|}{}                                                                        & $\hat {\sigma}_1$   & \multicolumn{1}{c|}{$\hat {\Pi}_{\left|+\right>}$}       & \multicolumn{1}{c|}{$2 \hat {\sigma}_1$}                           & \multicolumn{1}{c|}{$\sqrt{2} \hat {H}$}                           & $\hat {\sigma}_1 \cdot \hat {\Pi}_{\left|1\right>}$ \\ \cline{2-6} 
\multicolumn{1}{|c|}{}                                                                        & $\hat {\sigma}_3$   & \multicolumn{1}{c|}{$\hat {\Pi}_{\left|0\right>}$}       & \multicolumn{1}{c|}{$\sqrt{2} \hat {H}$}                           & \multicolumn{1}{c|}{$2 \hat {\sigma}_3$}                           & $\hat {\sigma}_3 \cdot \hat {\Pi}_{\left|+\right>}$ \\ \cline{2-6} 
\multicolumn{1}{|c|}{}                                                                        & $-i\hat {\sigma}_2$ & \multicolumn{1}{c|}{$\sqrt{2} \hat {\sigma}_1 \cdot \hat {H}$} & \multicolumn{1}{c|}{$\hat {\sigma}_1 \cdot \hat {\Pi}_{\left|0\right>}$} & \multicolumn{1}{c|}{$\hat {\sigma}_3 \cdot \hat {\Pi}_{\left|-\right>}$} & 0                                             \\ \hline
\end{tabular}
\caption{Table of the resulting Kraus operators given by the superposition of unitaries $\hat{U}_i$ and $\hat{U}_j$}
\label{tab:postexplained}
\end{table*}
There are three main reasons for the post-selected channel to take its form given in figure \ref{fig:post_selected_bloch_Plus}. Firstly, due to phases $s^{(a)}_2$ and $s^{(b)}_2$, the post-selection filters out the cases where both $N_a$ and $N_b$ implement $\hat \sigma_2$, which takes one eigenstates of the Hadamard to the other, as the random unitary. Secondly, when one of the channels implements $\hat \sigma_1$ and the other implements $\hat \sigma_3$, the resulting operator that an input state experiences is the Hadamard, which leaves the unitary's eigenstate unchanged. Thirdly, when one of the channels implements identity and the other implements either $\hat \sigma_1$ or $\hat \sigma_3$, the resulting operator is a projector onto the eigenstate with the positive eigenvalue for $\hat \sigma_1$ and $\hat \sigma_3$ respectively. These two states have a higher overlap to the eigenstate with the positive eigenvalue (than the negative eigenvalue) for the Hadamard unitary. A summary of these superpositions and the resulting operators can be found in table \ref{tab:postexplained}. Overall, this results in the post-selected channel leaving the positive-eigenvalued eigenstate for the Hadamard to be the least disturbed, resulting in a channel that has a Bloch sphere representation of an elliptical disc with its longest semi-axis in the direction of the Hadamard eigenstates and its center displaced towards the positive-eigenvalued Hadamard eigenstate, as illustrated in figure \ref{fig:post_selected_bloch}.

\section{Qutrit Tomography}\label{sec:qutrit_tomo}
Here, we perform  qutrit channel tomography for the aforementioned qutrit channels explicitly. To perform tomography on the vacuum state, we note that the unitary operators for the qutrit channels $N_a$ and $N_b$ can be described by
\begin{multline}\label{eq:two_qutrit}
    \hat{U}_{ij}^{(2T)}=\\\hat{U}_{i}^{(T0)}\otimes \hat{U}_{j}^{(T1)}=\\\left(\left|0\right>\left<0\right|^{(a)}\oplus \hat U_i^{(a)}\right) \otimes \left(\left|0\right>\left<0\right|^{(b)}\oplus \hat U_j^{(b)}\right),
\end{multline}
with $\left|0\right>\left<0\right|^{(a)}$ and $\left|0\right>\left<0\right|^{(b)}$ being projectors onto the zero photon states for channels $N_a$ and $N_b$ respectively, and $\hat{U}_{i}^{(T0)}$ and $\hat{U}_{j}^{(T1)}$ being the unitary operators for channels $N_a$ and $N_b$ described under the qutrit channel model. These sets of qutrit unitaries form the random unitary qutrit channels $\boldsymbol{\Phi}^{(T0)}$ and $\boldsymbol{\Phi}^{(T1)}$ (and that $\boldsymbol{\Phi}^{(2T)}=\boldsymbol{\Phi}^{(T0)}\otimes\boldsymbol{\Phi}^{(T1)}$), which have the feature that they preserve photon number.

For any accurate tomographic reconstruction of the channel $\boldsymbol{\Phi}$, we require a way to extract $\left|c\right|^2$ where
\begin{equation}
    \left|c\right|^2=\langle\psi'|\boldsymbol{\Phi}\left(|\psi\left\rangle\!\right\langle\psi|\right)|\psi'\rangle
\end{equation}
for a completely spanning set of states $|\psi\rangle$ $|\psi'\rangle$. In our experiment, we have direct access to the one photon subspace of the two-qutrit channel $\boldsymbol{\Phi}^{(2T)}$ in the form of our Mach-Zehnder channel $\boldsymbol{\Phi}^{(MZ)}$. In the Mach-Zehnder channel, the vacuum state for channel $N_a$ ($N_b$) can be accessed by preparing the path state to be in $\left|\textbf{b}\right \rangle$($\left|\textbf{a}\right \rangle$). 

The probability of measuring a certain output state $\left|\psi_{path}'\right \rangle\otimes\left|\psi_{pol}'\right \rangle$ given an input state $\left|\psi_{path}\right \rangle\otimes\left|\psi_{pol}\right \rangle$ is given by $\left|c_{MZ}\right|^2$, where
\begin{multline}\label{eq:data_eq}
    c_{MZ}=\left\langle\psi_{pol}'\right| \hat U_i^{(a)} \left|\psi_{pol}\right \rangle\cdot\left\langle\psi_{path}'|\textbf{a}\left>\!\right<\textbf{a}|\psi_{path}\right \rangle\\
    + \left\langle\psi_{pol}'\right| \hat U_i^{(b)} \left|\psi_{pol}\right \rangle\cdot\left\langle\psi_{path}'|\textbf{b}\left>\!\right<\textbf{b}|\psi_{path}\right \rangle.
\end{multline}
To perform qutrit channel tomography on channel $N_a$, we need to set $\hat U_j^{(b)}\left|\psi_{pol}\right>=\left|\psi_{pol}'\right>$, where $\hat U_j^{(b)}$ takes the input polarization state to the output polarization state. Thus, the probability amplitude $c_{T0}$ for the qutrit channel can be found by substituting this condition into equation \ref{eq:data_eq}. Performing the substitution, we have
\begin{multline}
    c_{T0}=\left\langle\psi_{pol}'\right| \hat U_i^{(a)} \left|\psi_{pol}\right \rangle\cdot\left\langle\psi_{path}'|\textbf{a}\left>\!\right<\textbf{a}|\psi_{path}\right \rangle\\
    + \left\langle\psi_{path}'|\textbf{b}\left>\!\right<\textbf{b}|\psi_{path}\right \rangle\\
    =\left\langle\psi_{trit}'\right|\hat U_i^{(T0)}\left|\psi_{trit}\right\rangle,
\end{multline}
where 
\begin{align}
\begin{split}
    \ket{\psi_{trit}}=a_H&\ket{\textbf{a}}\otimes\ket{H}\\
    +a_V&\ket{\textbf{a}}\otimes\ket{V}\\
    +a_{0}&\left|\textbf{b}\right\rangle\otimes\left|0\right\rangle
\end{split}
\end{align}
and 
\begin{align}
\begin{split}
    \ket{\psi_{trit}^{'}}=a_H^{'}&\ket{\textbf{a}}\otimes\ket{H}\\
    +a_V^{'}&\ket{\textbf{a}}\otimes\ket{V}\\
    +a_{0}^{'}&\left|\textbf{b}\right\rangle\otimes\left|0\right\rangle
\end{split}
\end{align}
where $a_{0}$ ($a'_{0}$), $a_{H}$ ($a'_{H}$), and $a_{V}$ ($a'_{V}$) is the probability amplitude of zero photons, one $\ket{H}$ photon, and one $\ket{V}$ photon being sent to (measured from) channel $N_a$. We also note that $\hat U_i^{(T0)}$ acts on the qutrit Hilbert space with basis vectors $\ket{\textbf{a}}\otimes\ket{H}$,  $\ket{\textbf{a}}\otimes\ket{V}$, and  $\ket{\textbf{b}}\otimes\ket{0}^{(a)}$. $a_{0}$ and $a'_{0}$ can take on the values of $1$ and $0$ when one of the paths of the interferometer is physically blocked, or the value of $1/\sqrt{2}$ when both paths are unblocked. The probability of measuring some output state given any input state will, as a result, be independent of polarization when the path qubit is in the $\left|\textbf{b}\right>$ state, effectively reducing the dimensions of the channel from a two-qubit channel to a qutrit channel where three orthogonal states exist for the entire apparatus -- one for the photon in path 1, one for the photon in path 0 and horizontally polarized, and one for the photon to be in path 0 and vertically polarized. This path $\left|\textbf{b}\right>$ state can thus be re-labelled as the vacuum state, as a reference to the fact that the photon is not in path 0. A similar procedure is repeated for channel $N_b$. 
\bibliographystyle{quantum}
\bibliography{suposechannel}
\end{document}